\documentclass[journal]{IEEEtran}
\usepackage{graphicx}
\usepackage{mathptmx}
\usepackage{amsmath}
\usepackage{amssymb}
\usepackage{theorem}
\usepackage{cite}
\usepackage{array}
\usepackage{url}

\centerfigcaptionstrue

\begin{document}

\title{Orthogonal Multiple Access with Correlated Sources: Achievable Region and Pragmatic Schemes}

\author{A.~Abrardo, G.~Ferrari, M.~Martal\`o, M.~Franceschini, and R.~Raheli
\thanks{A.~Abrardo is with the Department of Information Engineering, University of Siena, Italy. Email: abrardo@dii.unisi.it. G.~Ferrari, M.~Martal\`o and R.~Raheli are with the Department of Information Engineering, University of Parma, Italy. Email: \{gianluigi.ferrari,marco.martalo,raheli\}@unipr.it. M.~Franceschini is with IBM T.J. Watson Research Center, Yorktown Heights, NY, USA. Email: franceschini@us.ibm.com. This paper was presented in part at the 2009 and 2010 \emph{Information Theory and Applications Workshop} (ITA), UCSD, San Diego, CA, USA.}
}

\maketitle
\thispagestyle{empty}

\begin{abstract}
In this paper, we consider orthogonal multiple access coding schemes, where correlated sources are encoded in a distributed fashion and transmitted, through additive white Gaussian noise (AWGN) channels, to an access point (AP). At the AP, component decoders, associated with the source encoders, iteratively exchange soft information by taking into account the source correlation. The first goal of this paper is to investigate the ultimate achievable performance limits in terms of a multi-dimensional feasible region in the space of channel parameters, deriving insights on the impact of the number of sources. The second goal is the design of pragmatic schemes, where the sources use ``off-the-shelf'' channel codes. In order to analyze the performance of given coding schemes, we propose an extrinsic information transfer (EXIT)-based approach, which allows to determine the corresponding multi-dimensional feasible regions. On the basis of the proposed  analytical framework, the performance of pragmatic coded 
schemes, based on serially concatenated convolutional codes (SCCCs), is discussed.
\end{abstract}



\pagenumbering{arabic}

\section{Introduction} \label{sec:intro}
The efficient transmission of correlated signals, observed at different nodes, to one or more collectors is one of the main challenges in various networking scenarios, e.g., wireless sensor networks~\cite{AkSuSaCa02}. In the case of one collector node, this problem is often referred to as reach-back channel problem~\cite{BaSe06,GuKu00,Ga05}. In the case of separated additive white Gaussian noise (AWGN) channels, the separation between source (up to the Slepian-Wolf limit) and channel coding is known to be optimal~\cite{ShVe95,BaSe06}. However, implementing a practical system based on separation, i.e., given by distributed source coding (DSC) followed by channel encoding, is not straightforward~\cite{BaMi01,XiLiCh04} and the design of practically good codes is still an open issue~\cite{ZhEf03}. 

Alternative approaches are represented by cooperative source-channel coding and distributed joint source-channel coding (JSCC). In the JSCC case, no cooperation among sources is required, each source is independently encoded, and the correlation between the sources is exploited at the joint decoder by means of joint channel decoding (JCD)~\cite{GaZh01,DaLaMo05,DaLaMo06,GaZhZh07}. In other words, for a given source neither the data transmitted from the other sources nor the correlation model are available at the encoder. The correlation model between the sources must instead be assumed to be known at the (common) receiver, which aims at the reconstruction of the information streams transmitted by the sources. The problem of designing good codes for this scenario has been, however, only partially addressed. In~\cite{GaZhZh07}, the authors state that for two orthogonal channels the type of concatenated code utilized for the encoding process is not critical, and good results can be obtained, provided that 
powerful codes are employed. In~\cite{ZhAlBaMi04}, recursive non-systematic convolutional encoders are proposed as constituent encoders for heavily biased sources, leading to a signal-to-noise ratio (SNR) penalty between 0.74~dB and 1.17~dB with respect to the Shannon limit. In~\cite{YePfNa10}, optimized low-density parity-check (LDPC) codes are designed, by means of puncturing and proper iterative decoding schedule at the access point (AP). Extensions to universal codes (i.e., capacity-achieving codes for all possible channel parameters) through spatial coupling has been also recently considered~\cite{YePfNa10,YePfNa12,NgYePfNa12}. More precisely, the approaches in~\cite{YePfNa10,YePfNa12,NgYePfNa12}, relative to a two-source scenario, have the following characteristics: at each source, LDPC coding is used; at the AP, message-passing decoding is carried out on a joint bipartite graph (combining the graphs of the two codes) and the asymptotic performance, for infinite codeword length, is investigated. 
However, the extension of the proposed joint graph-based approach to an arbitrary number of sources is a challenging research direction. Another interesting approach has been presented in~\cite{AnMa12}, where practical concatenated coded schemes are designed for faded multiple-input multiple-output (MIMO) scenarios. However, the scheme is evaluated only for the case of two sources.

In this paper, we consider a generic number of correlated sources which transmit to a common AP through orthogonal AWGN channels. The sources do not explicitly use source codes, but only channel codes. At the AP, a proper iterative receiver is used to exploit the source correlation. This extends our previous works for two-source scenarios~\cite{AbFeMaFrRa09,MaFeAbFrRa10}, as well as~\cite{AbFeMa11}, where practical coding/decoding schemes have been designed in the presence of block faded channels. The first contribution of this paper is to shed light on the characterization of JCD schemes with an arbitrary number of correlated sources, by characterizing the multi-dimensional achievable region in the space of channel parameters for an arbitrary number of sources. It will be shown that a few characteristic points are sufficient to accurately characterize this achievable region. The asymptotic behavior, for a large number of sources, is also investigated. The impacts of the correlation level and of the 
number of sources, as well as the speed of convergence of the achievable region to the asymptotic limit, are discussed. To the best of our knowledge, this is the first work which considers scenarios with more than two sources. This is of interest, for instance, in wireless sensor networking, where many source nodes transmit their correlated data to a common AP.

On the basis of the characterization of the feasible region, as a second contribution of the paper, pragmatic schemes are discussed, where ``off-the-shelf'' channel codes are used at the sources, making the proposed iterative receiver directly scalable. In particular, we consider serially concatenated convolutional codes (SCCCs). By using density evolution, we propose an operational extrinsic information transfer (EXIT)-inspired approach to evaluate the performance, in terms of achievable region, of the given SCCCs with iterative decoding at the AP. Although optimal channel code design goes beyond the scope of this work, our characterization of the achievable region shows clearly 
that channel coding can be easily optimized in the presence of unbalanced channel conditions, i.e., where at least one rate is sufficiently low. In this unbalanced scenario, the beneficial impact of an increasing number of sources and, therefore, a more reliable a priori information for a given decoder, can be exploited. Our results also suggest that a code optimized for unbalanced rates tends, for increasing number of sources, to perform well also with balanced rates, provided it is suitably ``optimized.'' These considerations are justified by considering properly designed SCCCs.

This paper is structured as follows. In Section~\ref{sec:scenario}, preliminaries on the scenario of interest are given. In Section~\ref{sec:fr}, the multi-dimensional achievable region is introduced together with its information-theoretic asymptotic characterization. In Section~\ref{sec:jcd}, the principle of JCD is concisely reviewed. In Section~\ref{sec:exit}, an EXIT chart-based analysis is derived. In Section~\ref{sec:performance}, performance results relative to SCCCs are presented and discussed. Finally, Section~\ref{sec:concluding} concludes the paper.

In the following sections, the notation $p(\pmb{A})$ denotes the joint probability density function (PDF) of the continuous-value elements of a matrix $\pmb{A}$. Similarly, $P(\pmb{B})$ denotes the joint probability mass function (PMF) of the discrete-value elements of a matrix $\pmb{B}$.

\section{Scenario} \label{sec:scenario}
Consider $N$ spatially distributed nodes which sense, i.e., receive at their inputs, binary information sequences $\pmb{x}^{(\ell)} = (x_1^{(\ell)},\ldots,x_k^{(\ell)})^T$, where $\ell=1,\ldots,N$ denotes the node index, $k$ is the sequence length assumed equal for all sources, and $(\cdot)^T$ denotes the transpose operator. The information symbols of each sequence are assumed to be independent with $P(x_i^{(\ell)} =0)=P(x_i^{(\ell)} = 1) = 0.5$ and the following sequence correlation model is considered:
\begin{equation} \label{eq:corr_model}
x_i^{(\ell)} = b_i \oplus z_i^{(\ell)} \hspace{1cm} i=1,\ldots,k \hspace{1cm} \ell=1,\ldots,N
\end{equation}
where $\{b_i\}$ are independent and identically distributed (i.i.d.) binary random variables and $\{z_i^{(\ell)}\}$ are i.i.d. binary random variables with $P(z_i^{(\ell)}=0)=\rho$, with $1/2\leq\rho\leq1$. This correlation model corresponds to a scenario where the sources sense the output of a set of binary symmetric channels (BSCs), with cross-over probability $1-\rho$, whose input, at the $i$-th epoch, $i\in \{1,\ldots,k\}$, is a common information bit $b_i$. Obviously, if $\rho = 0.5$ there is no correlation between the binary information sequences $\{\pmb{x}^{(\ell)}\}_{\ell=1}^N$, whereas if $\rho = 1$ they are identical with probability 1. According to the chosen correlation model, the a priori joint PMF of the information sequences at the inputs of the $N$ nodes at the $i$-th epoch can be computed. By standard manipulations, one can show that
\begin{equation}
P(\pmb{x}_i) = \sum_{b_i=0,1} P(\pmb{x}_i|b_i)P(b_i) 
= \frac{1}{2}\underbrace{\left[\rho^{n_{\rm z}} (1-\rho)^{N-n_{\rm z}} + (1-\rho)^{n_{\rm z}} \rho^{N-n_{\rm z}}\right]}_{f(\rho,N,n_{\rm z})}
 \label{eq0AA}
\end{equation}
where $\pmb{x}_i=(x_i^{(1)},\ldots,x_i^{(N)})^T$ is the column vector denoting the bits at the input of the various nodes at time epoch $i$, $n_{\rm z}=n_{\rm z}(\pmb{x}_i)$ is the number of zeros in $\pmb{x}_i$, and the compact notation $f(\rho,N,n_{\rm z})$ has been introduced for later usage. The considered model may be representative of several communication scenarios. For example, it may model wireless sensor networks, where a set of nodes collect and transmit correlated data (e.g., they arise from the same physical phenomenon) to a common sink.

In Fig.~\ref{Schema}, the overall model for the multiple access scheme of interest is shown: $n$ source nodes communicate directly (and independently of each other) to the AP.
\begin{figure}
\begin{center}
\includegraphics[width=0.48\textwidth]{./images/schema.eps}
\end{center}
\caption{Proposed multiple access communication scenario: $N$ source nodes communicate directly to the AP.}
\label{Schema}
\end{figure}
The information sequence at the $\ell$-th source node is encoded using a binary linear code, denoted as $\mathcal{C}_\ell$ ($\ell=1,\ldots,N$) with codewords $\{\pmb{s}^{(\ell)}\}_{\ell=1}^N$ ($s_i^{(\ell)}\in\{0,1\}$, $i=1,\ldots,n$)---for simplicity, the codeword length $n$ is assumed equal for all source nodes. Therefore, the encoding rate at each source is $r=k/n$. The goal of the communication system is to recover, at the AP, the information signals $\{\pmb{x}^{(\ell)}\}_{\ell=1}^N$ with arbitrarily small probability of error. Assuming that binary phase shift keying (BPSK) is the used modulation format, after matched filtering and carrier-phase recovery, the real observable at the AP, relative to a transmitted binary information symbol, can be expressed as
\begin{equation}
y^{(\ell)}_i = \nu^{(\ell)}_i  + \eta^{(\ell)}_i = \sqrt{E_{\rm c}^{(\ell)}} \left(2s^{(\ell)}_i-1\right)  + \eta^{(\ell)}_i 
\hspace*{15mm} i=1,\ldots,n \qquad \ell=1,\ldots,N
 \label{eq1AA}
\end{equation}
where $\{\nu_i^{(\ell)}\}$ denote the antipodal transmitted BPSK symbols with energy $E_{\rm c}^{(\ell)}$ and $\{\eta^{(\ell)}_i\}$ are independent AWGN random variables with zero mean and variance $N_0/2$.

For conciseness, the following matrices are introduced:
\begin{eqnarray*}
\pmb{X} &\triangleq& \left(\pmb{x}_1, \pmb{x}_2, \ldots, \pmb{x}_k\right) = \left(\pmb{x}^{(1)},\pmb{x}^{(2)},\ldots, \pmb{x}^{(N)}\right)^T \\
\pmb{S} &\triangleq& \left(\pmb{s}_1, \pmb{s}_2, \ldots, \pmb{s}_n\right) = \left(\pmb{s}^{(1)},\pmb{s}^{(2)},\ldots, \pmb{s}^{(N)}\right)^T \\
\pmb{Y} &\triangleq& \left(\pmb{y}_1, \pmb{y}_2, \ldots, \pmb{y}_n\right) = \left(\pmb{y}^{(1)},\pmb{y}^{(2)},\ldots, \pmb{y}^{(N)}\right)^T .
\end{eqnarray*}
In other words, $\pmb{X}$ is an $N\times k$ matrix whose rows are the information bits at each source. Similarly, $\pmb{S}$ is an $N\times n$ matrix whose rows are the codewords transmitted by each source encoder and $\pmb{Y}$ is an $N\times n$ matrix whose rows are the received vectors at the output of each of the $N$ orthogonal channels.

\section{Achievable Region} \label{sec:fr}
\subsection{Characterization} \label{subsec:characterization_FR}
\begin{figure}
\centering
\includegraphics[width=0.48\textwidth]{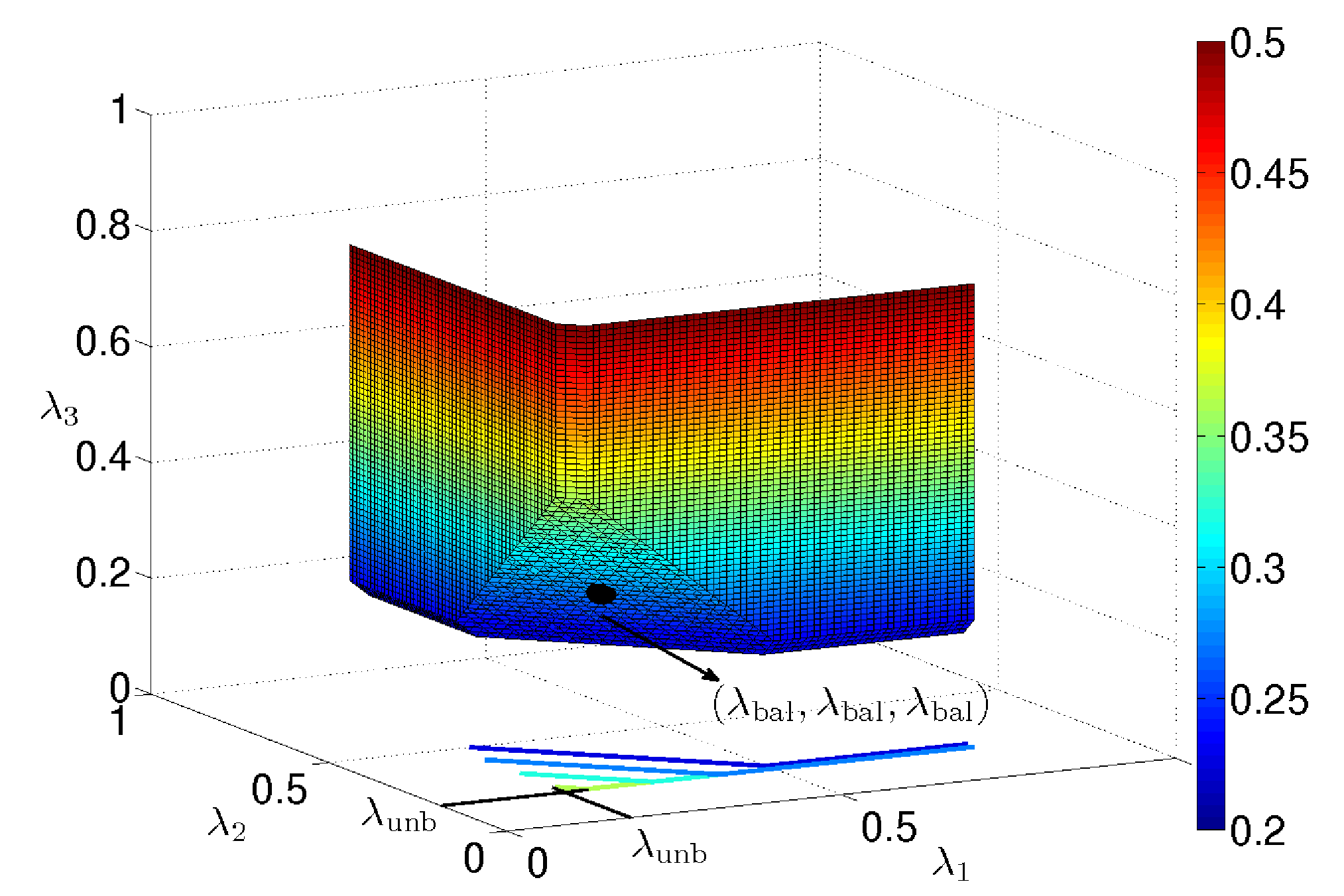}
\caption{Achievable region for $N=3$, $\rho=0.95$, and $r=1/2$ at each source.}
\label{img:feasible}
\end{figure}

In the described scenario, the performance achievable by a DSC scheme followed by channel coding is identical to that achievable if the sources were jointly channel encoded~\cite[Sec. 15.4]{CoTh06}. The Slepian-Wolf (SW) theorem allows to determine the achievable rate region for the case of separate lossless encoding of correlated sources. Denoting by $r_\ell^{\rm s}$ the source encoding rate for the $\ell$-th transmitter, the SW region~\cite[Sec. 15.4.3]{CoTh06} can be compactly formulated as the intersection of the family of inequalities
\begin{equation}
\sum\limits_{m=1}^{p} r_{\ell_m}^{\rm s} \ge H\left(N\right)-H\left(N-p\right)
 \label{eq3NN}
\end{equation}
where $p\in \{1,\ldots,N\}$, $\{\ell_1,\ldots,\ell_p\} \subseteq \{1,\ldots,N\}$, and
\begin{equation}
H(N) \triangleq -\frac{1}{2}\sum\limits_{n_{\rm z}=0}^{N} {N \choose n_{\rm z}}f(\rho,N,n_{\rm z}) \log_2\left\{\frac{1}{2}f(\rho,N,n_{\rm z})\right\}
\label{eq0AAbis}
\end{equation}
with the conventional assumption that $H(0) = 0$. The formulation (\ref{eq3NN})-(\ref{eq0AAbis}) can be derived by straightforward manipulations and can be found, e.g., in~\cite{AbFeMa11}. By assuming that source coding is followed by channel coding, the channel code rates $\{r_\ell^{\rm c}\}_{\ell=1}^N$ may be expressed as
\begin{equation}
r_\ell^{\rm c} = r_\ell^{\rm s} \cdot r
 \label{eq2NNb}
 \end{equation}
where we recall that $r=k/n$. The channel code rates must satisfy the following Shannon bounds:
\begin{equation}
r_\ell^{\rm c} \le \lambda_{\ell}  \hspace{7mm} \ell=1,\ldots,N
 \label{eq3NNb}
\end{equation}
where $\lambda_{\ell}$ is the capacity\footnote{The specific expression of $\lambda_\ell$ should take into account possible input constraints, such as the modulation format BPSK used in Section~\ref{sec:performance}.} (dimension: [bits per channel use]) at the AP, relative to the $\ell$-th link with SNR equal to $\gamma_{\ell}$~\cite{Pr01}.  As noted in Section~\ref{sec:intro}, compressing each source up to the SW limit and then utilizing independent capacity-achieving channel codes allows to achieve the ultimate performance limits~\cite{ShVe95,BaSe06}. Combining (\ref{eq3NN}), (\ref{eq2NNb}), and (\ref{eq3NNb}), an achievable region of individual capacity values characterizing the set of orthogonal channels can be identified by the following inequalities to be jointly satisfied by the link capacities $\{\lambda_{\ell}\}_{\ell=1}^N$:
\begin{equation}
\begin{array}{c}
\sum\limits_{m=1}^{p} \lambda_{\ell_m} \ge r \, \left[H\left(N\right)-H\left(N-p\right)\right]\\
\end{array}
 \label{eq5NN}
\end{equation}
for $p \in \{1,\ldots,N\}$ and $ \{\ell_1,\ldots,\ell_p\} \subseteq  \{1,\ldots,N\}$. For conciseness, we refer to this region as \emph{achievable region}. From a geometric point of view, the achievable region defines a polymatroid structure and its border corresponds to a non-closed convex $N$-dimensional polytope~\cite{Ha79}. In general, the border of the achievable region is given by the intersection of the $2^N-1$ hyperplanes defined by (\ref{eq5NN}). Fig.~\ref{img:feasible} depicts a typical achievable region for $N=3$, $\rho=0.95$, and $r=1/2$.

A few characteristic points can be identified, for each finite value of $N$, on the border of the achievable region. In particular, two types of characteristic operational regions, denoted as ``balanced'' and ``unbalanced,'' are of interest. The \emph{balanced} case refers to the characteristic point, on the border of the achievable region, corresponding to a scenario where all sources are transmitted at a rate equal to the same single-channel capacity, i.e., $\lambda_1 = \lambda_2 = \cdots = \lambda_N$. This common value, denoted as $\lambda_{\rm bal}$, can be determined by considering the hyperplane associated with $p=N$ in (\ref{eq5NN}), thus obtaining
$$
\sum_{i=1}^N \lambda_i = N\lambda_{\rm bal} = r\, H(N)
$$
and, therefore,
\begin{equation} \label{eq:lbal}
\lambda_{\rm bal} \triangleq r\frac{H(N)}{N} .
\end{equation}
The \emph{unbalanced} case, instead, refers to the portion of the achievable region characterized as follows: $N-1$ sources, e.g., sources from 1 to $N-1$, are associated with values of $\lambda_i$ ($i=1,\ldots,N-1$) sufficiently large to satisfy the corresponding constraints of type (\ref{eq5NN}). In this case, $\lambda_{\rm unb}$ is the smallest value of $\lambda_N$ such that the operational point lies on the border of the achievable region. This corresponds to considering the hyperplane associated with $p=1$ and $\ell_1=N$ in (\ref{eq5NN}), thus obtaining
\begin{equation} \label{eq:lunb}
\lambda_N = r \left[H(N)-H(N-1)\right] \triangleq \lambda_{\rm unb} .
\end{equation}
Note that $\lambda_{\rm bal}$ and $\lambda_{\rm unb}$ are functions of $N$ but, for the sake of readability, we will not explicitly indicate the dependence on $N$---the context will eliminate any ambiguity. For a given value of $N$, unlike the unique characteristic \emph{point} associated with $\lambda_{\rm bal}$, there are infinite operational points associated with $\lambda_{\rm unb}$. Fig.~\ref{img:feasible} shows the achievable region for $N=3$, along with the characteristic values $\lambda_{\rm bal}$ and $\lambda_{\rm unb}$ (associated with 3 hyperplanes on the border).

\begin{figure}
\begin{center}
\includegraphics[width=0.48\textwidth]{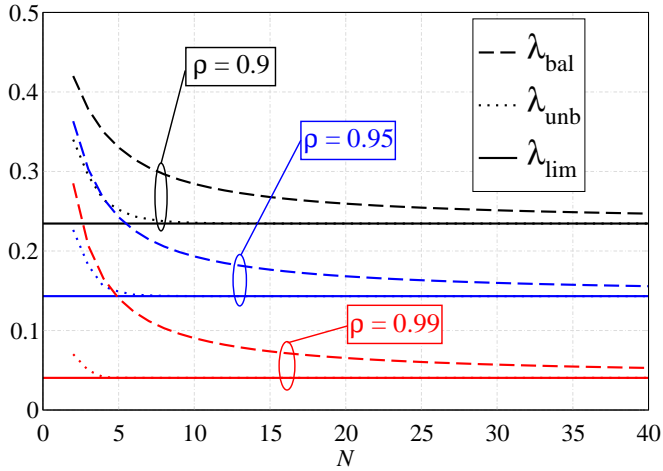}
\caption{$\lambda_{\rm bal}$, $\lambda_{\rm unb}$, and $\lambda_{\rm lim}$, as functions of $N$, in a scenario with $r=1/2$. Three different values for $\rho$ are considered: (i) 0.9, (ii) 0.95, and (iii) 0.99.}%
\label{img:limits_FR}
\end{center}
\end{figure}

We now investigate the behavior of the characteristic values $\lambda_{\rm bal}$ and $\lambda_{\rm unb}$. To this end, the information sequence at the $\ell$-th node and $i$-th epoch can be viewed as a stationary stochastic process in the index $\ell$ (for fixed $i$). Due to stationarity, the following facts can be observed~\cite[Ch.~4]{CoTh06}:
\begin{eqnarray}
&& \lambda_{\rm bal} \geq \lambda_{\rm unb} \qquad \forall N \nonumber \\
&&\lim_{N\rightarrow +\infty} \lambda_{\rm unb} = \lim_{N\rightarrow +\infty} \lambda_{\rm bal} \triangleq \lambda_{\rm lim} = r H_{\rm b}(\rho) \label{eq:inf_rate}
\end{eqnarray}
where $H_{\rm b}(\rho)$ is the entropy rate of the binary stochastic process $\{x_i^{(\ell)}\}$. In Fig.~\ref{img:limits_FR}, $\lambda_{\rm bal}$, $\lambda_{\rm unb}$, and $\lambda_{\rm lim}$ are shown, as functions of $N$, in a scenario with $r=1/2$ and three different values of $\rho$: (i) 0.9, (ii) 0.95, and (iii) 0.99.
First, one can observe that, for increasing values of $n$, $\lambda_{\rm bal}$ and $\lambda_{\rm unb}$ are decreasing functions of $N$. Therefore, the widest projection of the border of the achievable region on a two-dimensional plane (e.g., the $(\lambda_1,\lambda_2)$ plane) enlarges for increasing values of $N$. Moreover, it can be also verified that $\lambda_{\rm bal}\geq\lambda_{\rm unb}\,\, \forall N$, and that both $\lambda_{\rm bal}$ and $\lambda_{\rm unb}$ approach the same asymptotic value $\lambda_{\rm lim}$. We can also observe that in the unbalanced case the convergence is significantly faster, especially for increasing values of $\rho$. Therefore, the shape of the $N$-dimensional achievable region tends, for increasing values of the number of sources, to that of a translated hyperoctant defined by the following set of inequalities:
$$
\lambda_i\geq\lambda_{\rm lim} \hspace*{10mm} i=1,2,\ldots,N .
$$
In fact, the asymptotic achievable region would not be a translated hyperoctant only if a point in the hyperplane associated with $p=N$ did not have the same limit. However, the achievable region is a polymatroid defined by the hyperplanes in (\ref{eq3NN}) and this is not allowed.

\subsection{Speed of Convergence} \label{subsec:speed}
We now analyze the speed of convergence of the considered multiple access schemes in terms of how many sources are needed to achieve the asymptotic performance associated with a very large value of $N$. On the basis of the observations carried out at the end of the previous subsection, the convergence speed can be interpreted as the speed at which the border of the achievable region tends to adhere to the border of the asymptotic translated hyperoctant ($N\rightarrow+\infty$).

\begin{figure*}
\centering
\begin{tabular}{ccc}
\hspace*{-5mm}
\includegraphics[clip,width=0.32\textwidth]{images/feasible1.eps} &
\includegraphics[clip,width=0.32\textwidth]{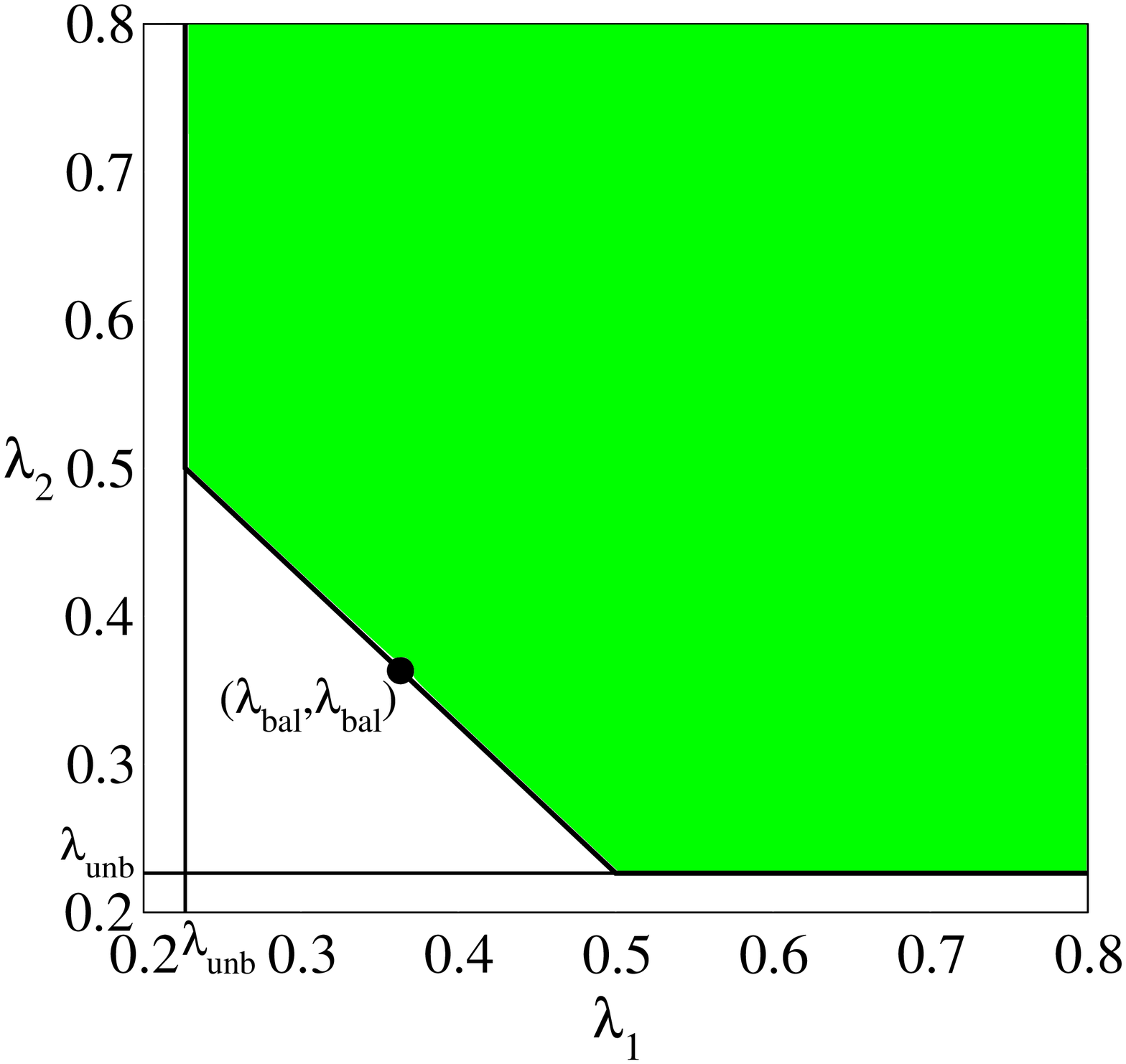} &
\includegraphics[clip,width=0.32\textwidth]{images/feasible3.eps}\\
(a) & (b) & (c)
\end{tabular}
\caption{Projections of the achievable region for various values of $N$ and $\rho$: (a) 0.9, (b) 0.95, and (c) 0.99.}
\label{fig:feasible}
\end{figure*}

In Fig.~\ref{fig:feasible}, the $(\lambda_1,\lambda_2)$ projections (solid lines) of the achievable region are shown for various values of $N$ and $\rho$ equal to: (a) 0.9, (b) 0.95, and (c) 0.99. For each projection contour associated with each finite value of $N$, the dashed lines indicate, in the two-dimensional projection plane, the ``missing'' triangle with respect to the projection of the corresponding translated hyperoctant.
As one can observe from Fig.~\ref{fig:feasible}, for $N=6$ the projection of the achievable region is very close to the asymptotic achievable region (i.e., that for $N\rightarrow\infty$) predicted by our analytical framework. In particular, the borders of the achievable region in the unbalanced zones approach very quickly the corresponding borders of the asymptotic translated hyperoctant. Even though convergence is slower in the balanced zone, from Fig.~\ref{fig:feasible} one may note that most of the gap from the asymptotic translated hyperoctant is ``filled.''

In order to compare the scenarios associated with different values of $N$ (i.e., different dimensionality), we consider, as a convergence indicator, for given values of $N$ and $\rho$, the area of the triangle identified by the dashed (horizontal and vertical) lines and the solid (diagonal) line. This area is denoted as $A(N,\rho)$ and two illustrative cases are shown in Fig.~\ref{fig:feasible}~(a) ($A(2,0.9)$) and in Fig.~\ref{fig:feasible}~(c) ($A(6,0.99)$). The rationale behind this choice is the fact that this area tends to zero for $N\rightarrow\infty$. Using straightforward geometric considerations, one can write
$$
A(N,\rho) = \frac{(2\lambda_{\rm bal}-\lambda_{\rm unb}-\lambda_{\rm unb})^2}{2} = 2(\lambda_{\rm bal}-\lambda_{\rm unb})^2.
$$

Since the (missing) area $A(N,\rho)$ is asymptotically a decreasing function of $N$, we can characterize the convergence in terms of its rate of reduction, for increasing values of $N$. In particular, for a given value of $N$, we introduce the relative area reduction with respect to the case with $N=2$, defined as follows:
$$
\chi(N,\rho) \triangleq \frac{A(N,\rho)}{A(2,\rho)} .
$$
We will refer to $\chi(N,\rho)$ as ``area ratio.''

\begin{figure}
\centering
\begin{tabular}{c}
\includegraphics[width=0.48\textwidth]{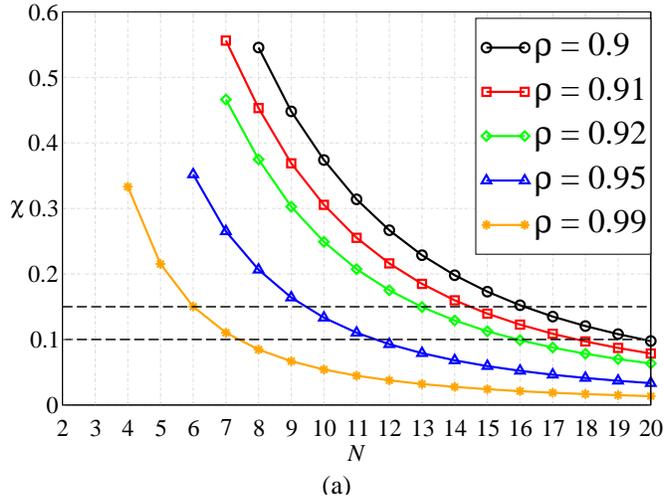} \\
(a)\\
\includegraphics[width=0.48\textwidth]{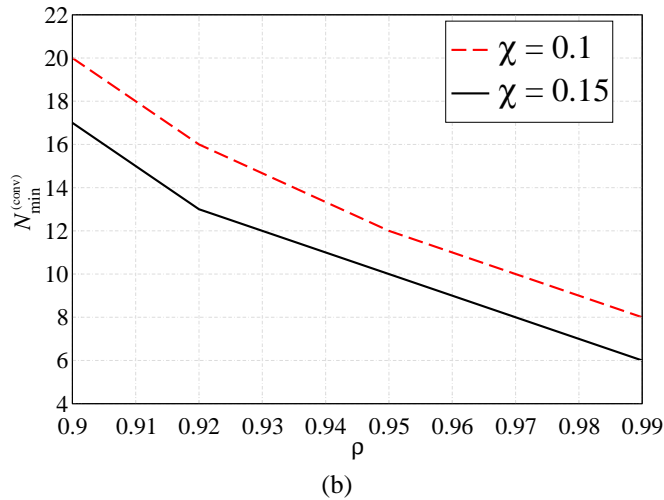} \\
(b)
\end{tabular}
\caption{(a) The area ratio $\chi$, as a function of $N$, for various values of $\rho$ and (b) the minimum value of $N$, as a function of $\rho$, to achieve a given convergence level (area reduction).}
\label{fig:norm_area}
\end{figure}

In Fig.~\ref{fig:norm_area}~(a), the area ratio is shown, as a function of $N$, for various values of $\rho$. For each value of $\rho$, the minimum considered value of $N$ is the one which practically guarantees convergence in the unbalanced zone. In particular, for each value of $\rho$, practical convergence is obtained if the difference $\lambda_{\rm unb}-\lambda_{\rm lim}$ is at most 1\% of the value of $\lambda_{\rm lim}$.
From the results in Fig.~\ref{fig:norm_area}~(a), one can derive the minimum number of sources needed to achieve a given area ratio, i.e., a given convergence level. Assuming that convergence is achieved when the area ratio reduces below a desired value (i.e., there is a desired area reduction), the corresponding minimum value of $N$, denoted as $N_{\min}^{\rm (conv)}$, can be determined. In Fig.~\ref{fig:norm_area}~(b), $N_{\min}^{\rm (conv)}$ is shown, as a function of $\rho$, for various values of $\chi$, also highlighted by the dashed horizontal lines in Fig.~\ref{fig:norm_area}~(a). The obtained results quantify the intuitive fact that the minimum number of sources required to achieve the desired convergence level reduces for increasing values of $\rho$. In particular, for $\rho\geq0.92$, our results show that $N_{\min}^{\rm (conv)}$ can be accurately approximated as a linearly decreasing function of $\rho$.

\section{JCD Principle} \label{sec:jcd}
As the ultimate performance limits (in terms of achievable region) have been characterized, it is of interest to understand how given (pragmatic) channel coding schemes perform. To this end, in the remainder of this paper we first recall the JCD principle and then generalize the EXIT chart-based method, introduced in~\cite{AbFeMaFrRa09,MaFeAbFrRa10}, to the performance analysis of channel codes in the multiple access scenario of interest with an arbitrary number of nodes.

Using the matrix notation introduced in Section~\ref{sec:scenario}, the joint maximum a posteriori probability (MAP) decoding rule, given that $\pmb{Y}$ is received, reads:
\begin{equation}
\hat{x}_i^{(\ell)} = \mathop{\rm argmax}\limits_{{x_i^{(\ell)}} = 0,1}
\sum\limits_{\hspace*{2mm}\pmb{X}\sim {x_i^{(\ell)}}} p\left(\pmb{Y}|\pmb{X}\right)P\left(\pmb{X}\right)
\label{eq1mp}
\end{equation}
where $i=1,\ldots,k$, $\ell=1,\ldots,N$, and the notation $\pmb{X}\sim {x_i^{(\ell)}}$ denotes that the summation runs over all variables in $\pmb{X}$ except $x_i^{(\ell)}$. From (\ref{eq1mp}), using standard manipulations one can write:
\begin{eqnarray}
\hat{x}_i^{(\ell)} &=& \mathop{\rm argmax}\limits_{{x_i^{(\ell)}} = 0,1}\!\!\!\!\!\sum\limits_{\hspace*{2mm}\pmb{X}\sim {x_i^{(\ell)}}}\!\!\!\!\! p\left(\pmb{Y}|\pmb{S}\right)P\left(\pmb{S}|\pmb{X}\right)P\left(\pmb{X}\right) \nonumber \\
&& \hspace*{-15mm}=\mathop{\rm argmax}\limits_{{x_i^{(\ell)}} = 0,1}\!\!\!\!\!
\sum\limits_{\hspace*{2mm}\pmb{X}\sim {x_i^{(\ell)}}} \prod_{\ell=1}^N P\left(\pmb{s}^{(\ell)}|\pmb{x}^{(\ell)}\right)\!\!\prod_{i=1}^{n} p\left(y_i^{(\ell)}|s_i^{(\ell)}\right)\!\! \prod_{i=1}^{k} P\left(\pmb{x}_i\right) \label{eq2mp}
\end{eqnarray}
where we have used the facts that the information sequences are coded independently and sent over orthogonal AWGN channels. The probability $P(\pmb{s}^{(\ell)}|\pmb{x}^{(\ell)})$ is equal to 1 if $\pmb{s}^{(\ell)}$ is the codeword associated with $\pmb{x}^{(\ell)}$ and 0 otherwise.

Equation (\ref{eq2mp}) admits a Tanner graph representation and a corresponding belief propagation (BP) solution, provided that $P(\pmb{s}^{(\ell)}|\pmb{x}^{(\ell)})$ ($\ell=1,\ldots,N$) can be expressed as a product of factors which depend on restricted subsets of all symbol variables. This is always possible if $\mathcal{C}_\ell$ ($\ell=1,\ldots,N$) are convolutional codes or a serial or parallel concatenation of convolutional codes (i.e., turbo codes). Another situation where equation (\ref{eq2mp}) easily admits a Tanner graph-based representation is when $\mathcal{C}_\ell$ are LDPC systematic codes.

\begin{figure}
\begin{center}
\includegraphics[width=0.48\textwidth]{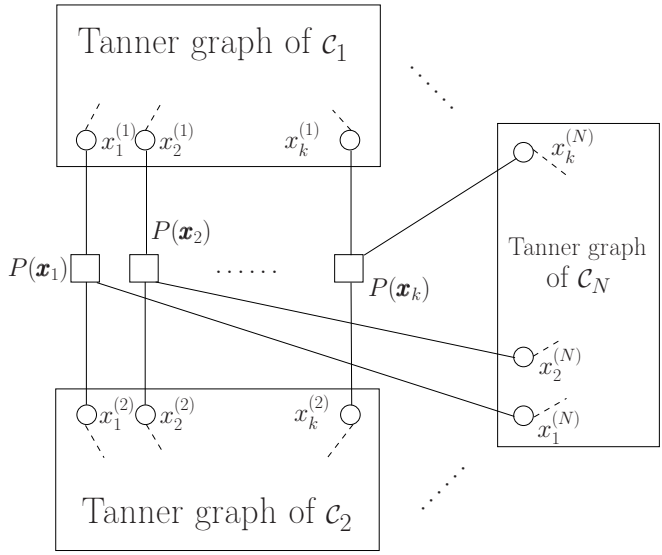}
\caption{Tanner graph for equation (\ref{eq2mp}).}
\label{F1}
\end{center}
\end{figure}

Consider $N$ separate Tanner graphs corresponding to the codes $\{\mathcal{C}_\ell\}_{\ell=1}^N$. A pictorial description of the global Tanner graph is shown in Fig.~\ref{F1}, where, for clarity, the variable nodes $\{\pmb{x}_i\}_{i=1}^k$ are explicitly shown.
Each single variable node $x_j^{(\ell)}$ ($j=1,\ldots,k$, $\ell=1,\ldots,N$) of the Tanner graph of $\mathcal{C}_\ell$ is connected to the corresponding node $x_j^{(m)}$ ($j=1,\ldots,k$, $m\neq \ell$) of the Tanner graph of $\mathcal{C}_{m}$ through a connection node, marked by the joint PMF $P(\pmb{x}_j)$. Note that this PMF depends on $\rho$. The connection nodes, upon receiving the logarithmic likelihood ratios (LLRs) messages from $N-1$ of the $N$ component Tanner graphs, send input LLRs to the other Tanner graph. The LLR output by the connection node for the $\ell$-th component Tanner graph at the $j$-th position, denoted as ${\rm LLR}_{{\rm out,}j}^{(\ell)}$, can be expressed as
\begin{equation}
{\rm LLR}_{{\rm out,}j}^{(\ell)} = \ln\, \, \frac{P\left(x_j^{(\ell)}=0\right)}{P\left(x_j^{(\ell)}=1\right)} 
= \ln\, \, \frac{\sum\limits_{\pmb{x}'_j}P\left(x_j^{(\ell)}=0|\pmb{x}'_j\right)P\left(\pmb{x}'_j\right)}{\sum\limits_{\pmb{x}'_j}P\left(x_j^{(\ell)}=1|\pmb{x}'_j\right)P\left(\pmb{x}'_j\right)} \label{eqn:n2}
\end{equation}
where $j=1,\ldots,k$, $\ln$ denotes the natural logarithm, and $\pmb{x}'_j = \pmb{x}_j\setminus x_j^{(\ell)}$ is the column vector denoting the bits at the input of the various nodes, with the exception of the $\ell$-th one, at time epoch $j$. The factors $\{P(\pmb{x}'_j)\}$ denote the probabilities coming from the other $N-1$ decoders (corresponding to the other sources). Assuming these $N-1$ outputs are independent, one can write
$$
P\left(\pmb{x}'_j\right) = \prod\limits_{\substack{m=1\\m\neq \ell}}^N P\left(x_j^{(m)}\right)
= \prod\limits_{\substack{m=1\\m\neq \ell}}^N\frac{e^{\overline{x}_j^{(m)}{\rm LLR}_{{\rm in,}j}^{(m)}}}{1+e^{{\rm LLR}_{{\rm in,}j}^{(m)}}}
$$
where ${\rm LLR}_{{\rm in,}j}^{(m)}$ is the LLR associated with the $j$-th bit coming from the $m$-th decoder and $\overline{x}_j^{(m)}$ is the logical negation of ${x}_j^{(m)}$. Note that ${\rm LLR}_{{\rm in,}j}^{(m)}$ ($m=1,\ldots,N$; $m\neq \ell$) may be seen as a priori information on the transmitted bits and can thus be easily taken into account by standard soft-input soft-output decoders. 

The conditional a posteriori probabilities $P(x_j^{(\ell)}=0|\pmb{x}'_j)$ and $P(x_j^{(\ell)}=1|\pmb{x}'_j)$ in \eqref{eqn:n2} can be computed by relying on the statistical characterization of the random variables $\{z_j^{(\ell)}\}$ given at the beginning of Section~\ref{sec:scenario}. After straightforward manipulations, one can write
\begin{equation} \label{eqn:n8}
{\rm LLR}_{{\rm out,}j}^{(\ell)} = \ln\frac{\sum\limits_{\pmb{x}'_j}f(\rho,N,n'_{\rm z}+1)\prod\limits_{\substack{m=1\\m\neq \ell}}^N\frac{e^{\overline{x}_j^{(m)}{\rm LLR}_{{\rm in,}j}^{(m)}}}{1+e^{{\rm LLR}_{{\rm in,}j}^{(m)}}}}
{\sum\limits_{\pmb{x}'_j}f(\rho,N,n'_{\rm z})\prod\limits_{\substack{m=1\\m\neq \ell}}^N\frac{e^{\overline{x}_j^{(m)}{\rm LLR}_{{\rm in,}j}^{(m)}}}{1+e^{{\rm LLR}_{{\rm in,}j}^{(m)}}}} .
\end{equation}
where $n'_{\rm z}$ is the number of zeros in the sequence $\pmb{x}'_j$. In~\cite{AbFeMa11}, a simplified sub-optimal version of (\ref{eqn:n8}), which takes into account pairwise a priori probabilities only, can be found. It is worth noting that the optimal combination rule (\ref{eqn:n8}) and its sub-optimal (pairwise) version in~\cite{AbFeMa11} coincide for $N=2$. Therefore, (\ref{eqn:n8}) reduces to equation (10) in~\cite{MaFeAbFrRa10} for $N=2$.\footnote{Note that (10) in~\cite{MaFeAbFrRa10} contains a typographical error. In fact, $e^{{\rm LLR}_{x,j}}$ and $e^{-{\rm LLR}_{x,j}}$ should be replaced by $e^{{\rm LLR}_{x,j}}/(1+e^{{\rm LLR}_{x,j}})$ and $1/(1+e^{{\rm LLR}_{x,j}})$, respectively.} Note also that the LLR transformation (\ref{eqn:n8}) is monotonic with respect to each input variable $\{{\rm LLR}_{{\rm in,}j}^{(m)}\}$.

The scheduling of the BP procedure on the overall graph can be serially performed as follows. The messages emitted by the function nodes $\{P(\pmb{x}_j)\}_{j=1}^{k}$ can be initialized to zero and ``internal'' BP iterations within the component Tanner graph $\mathcal{C}_1$ run. At the end of these BP iterations, the messages $\{{\rm LLR}_{{\rm in,}j}^{(\ell)}\}_{j=1}^{k}$ are fed to the connecting nodes $\{P(\pmb{x}_j)\}_{j=1}^{k}$ which, in turn, emit new LLRs for the other component Tanner graph $\mathcal{C}_2$. This operation is repeated for the computation of the LLRs to be fed into $\mathcal{C}_3$, and so on until $\mathcal{C}_N$. The iterations between the $N$ Tanner graphs, through the connection nodes, are referred to as ``external.'' Note that the results in Section~\ref{sec:performance} are obtained using this BP scheduling. However, different scheduling can be considered, leading to slightly different performance. As an example, in~\cite{YePfNa10} the BP procedure is performed, for $N=2$, on the 
overall Tanner graph, without resorting to internal and external iterations. Extending this approach to a scenario with $N\geq3$ is an open and challenging issue.

\section{Exit Chart-based Analysis} \label{sec:exit}
In order to evaluate the performance of the overall joint decoder, we consider an EXIT chart-based approach. In particular, we build upon the EXIT chart-based approach proposed in~\cite{DiDoPo01} and further analyzed in~\cite{AbFeMaFrRa09,MaFeAbFrRa10}, for the two-source scenario, to characterize the evolution of the LLRs within each component decoder. An extension of this EXIT chart-based analysis method to generic scenarios with $N\geq 2$ is presented in the remainder of this section. As mentioned in Section~\ref{sec:intro}, in this paper we will focus on component SCCCs, as this will allow to provide simple guidelines to select ``off-the-shelf'' SCCCs. We remark that the proposed framework can be applied also to a scenario where LDPC component codes (and decoders) are used~\cite{AbFeMaFrRa09,MaFeAbFrRa10}. However, it is difficult to provide simple guidelines for the selection of standard LDPC codes. Proper LDPC code design for $N\geq3$, as shown in~\cite{YePfNa10,YePfNa12} for $N=2$, is an interesting 
research direction which goes beyond the scope of this paper.

Without loss of generality, we focus on code $\mathcal{C}_\ell$, $\ell=1,\ldots,N$, and assume that the corresponding source transmits the all-zero sequence. Therefore, the corresponding decoder receives, at its input, a sequence of Gaussian observables specified by the channel SNR $\gamma_\ell$. The channel LLRs are fed to the inputs of the variable nodes. Density consistency is imposed by modeling the LLR pdf\footnote{The variable $z$ should not be confused with the output of the BSCs in the correlation model (\ref{eq:corr_model}).} $\Gamma_{\rm ch}(z)$ as Gaussian with mean $\mu_{\rm ch}$ and variance $2\mu_{\rm ch}$~\cite{RiUr01}. Accordingly, $\gamma_\ell = \mu_{\rm ch}^2/2\mu_{\rm ch} = \mu_{\rm ch}/2$. Using this assumption, the EXIT chart-based approach proposed in~\cite{DiDoPo01} allows to evaluate the SNR of the extrinsic information messages at the output of the component decoder.

\begin{figure}
\begin{center}
\includegraphics[width=0.48\textwidth]{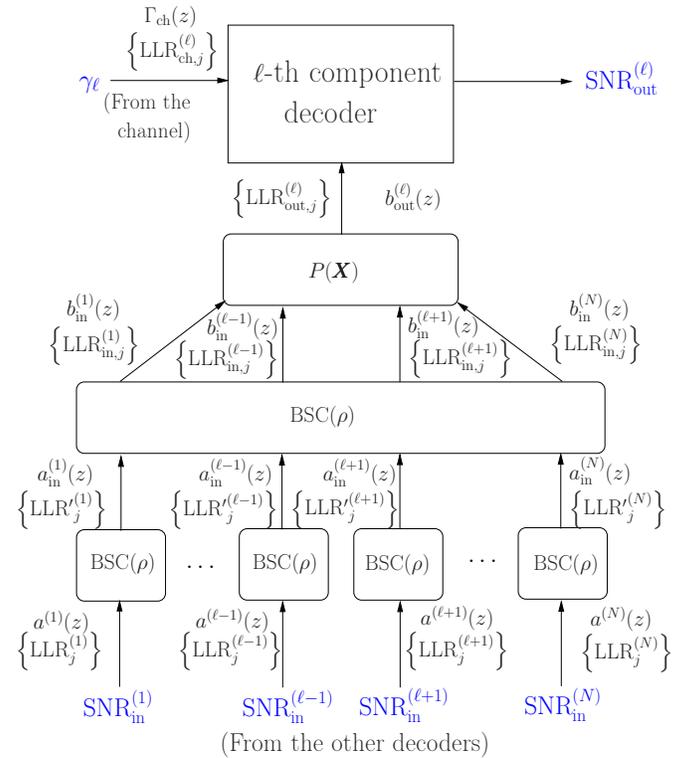}
\caption{Scheme for the analysis of the evolution of the a priori information.}%
\label{fig:DE}
\end{center}
\end{figure}

In Fig.~\ref{fig:DE}, an illustrative scheme to analyze the evolution of the a priori information through the $\ell$-th component decoder ($\ell=1,\ldots,N$), taking into account the soft information generated by the other decoders, is shown.
To account for the presence of a priori information coming from the other component decoders, let us denote by ${\textrm{SNR}}_{\textrm{in}}^{(m)}$ ($m=1,\ldots,N$, $m\neq \ell$) the SNR of external messages $\{{\rm LLR}_{j}^{(m)}\}_{j=1}^k$ entering the set of connection nodes characterized by the joint PMF $P(\pmb{X})$. We assume that also the messages $\{{\rm LLR}_{j}^{(m)}\}_{j=1}^k$ have a Gaussian distribution with mean $2 \,{\textrm{SNR}}_{\textrm{in}}^{(m)}$ and standard deviation$\sqrt{4\, {\textrm{SNR}}_{\textrm{in}}^{(m)}}$, so that each PDF is completely determined by the single parameter ${\textrm{SNR}}_{\textrm{in}}^{(m)}$. These messages are processed by the set of connection nodes with joint PMF $P(\pmb{X})$ to produce a priori information messages $\{{\rm LLR}_{{\rm out},j}^{(\ell)}\}_{j=1}^k$ for the variable nodes of the Tanner graph of $\mathcal{C}_\ell$.

Denote the PDFs of the messages $\{{\rm LLR}_{j}^{(m)}\}_{j=1}^k$ as $a^{(m)}(z)$ ($m=1,\ldots,N$; $m\neq\ell$) and the PDF of $\{{\rm LLR}_{{\rm out},j}^{(\ell)}\}_{j=1}^k$ as $b_{\rm out}^{(\ell)}(z)$. It is worth noting that an EXIT chart-based approach requires that all $\{a^{(m)}(z)\}_{m=1}^N$ are densities of messages corresponding to all-zero transmitted information sequences. Hence, taking into account the correlation model of $\{\pmb{x}^{(m)}\}_{m=1}^N$ introduced in Section~\ref{sec:scenario}  and based on a common virtual originating bit, for the purpose of analysis it is necessary to introduce two consecutive BSC-like blocks, each with cross-over probability $\rho$, for each sequence $\pmb{x}^{(\ell)}$, $\ell\neq m$. Since a BSC with parameter $\rho$ ``flips'' a bit at its input with probability $\rho$, the BSC-like block flips the sign of an input LLR with the same probability. The second BSC-like block is common for the set of the first BSC-like blocks, in the sense that it flips all 
outputs of the first BSC-like blocks exactly in the same way. At the output of the first block there is an estimate of the sequence $\{b_\ell\}$, whereas at the output of the second block there is an estimate of $\pmb{x}_\ell$. We remark that each of the first BSC-like blocks is associated with each of the other $N-1$ decoders, whereas the second (common) BSC-like block is the same for all soft messages at the output of the previous $N-1$ BSC-like blocks. This is compliant with the correlation model (\ref{eq:corr_model}).

The messages  at the input of the second BSC-like block (i.e., at the output of the first set of BSC-like blocks), denoted as $\{{\rm LLR'}^{(m)}_j\}$, are then characterized by the following PDFs:
\begin{equation} \label{eq:message_BSC}
a_{\rm in}^{(m)}(z) = \rho a^{(m)}(z) + (1-\rho) a^{(m)}(-z) \hspace*{5mm} m=1,\ldots,N; \,  m\neq \ell .
\end{equation}
The messages at the input of the set of connection nodes are denoted as $\{{\rm LLR}^{(m)}_{{\rm in},j}\}_{j=1}^k$. Note that this scheme is compliant with that, relative to a scenario with $N=2$, proposed in~\cite{AbFeMaFrRa09,MaFeAbFrRa10}: in fact, for $N=2$,the cascade of two BSC-like blocks with parameter $\rho$ is equivalent to a single BSC-like block with parameter $p_{\rm f}=\rho^2+(1-\rho)^2$. We remark that this block combination can not be extended to $N>2$, as the second BSC-like block is common for the set of the first BSC-like blocks.

The PDF $b_{\rm out}^{(\ell)}(z)$ of $\{{\rm LLR}_{{\rm out},j}^{(\ell)}\}_{j=1}^k$ can eventually be computed according to (\ref{eqn:n8}), with input messages $\{{\rm LLR}_{{\rm in},j}^{(m)}\}_{j=1}^k$, by applying well-known results for PDF transformation~\cite{Pa91}. Note that, unlike $a^{(m)}(z)$ and $\Gamma_{\rm ch}^{(\ell)}$, $b_{\rm in}^{(m)}(z)$ cannot be Gaussian. It can be verified that the analytical computation of $b_{\rm out}^{(\ell)}(z)$ has an exponential complexity on the order of $N_{\rm PDF}^{N-1}$, where $N_{\rm PDF}$ is the number of samples of the numerical representation of the PDFs used in the computer solver. In fact, for each output sample in $b_{\rm out}^{(\ell)}(z)$, all possible combinations (of length $N-1$) from the input PDFs $\{a^{(\ell)}(z)\}$ which return this sample should be analyzed. In order to limit the computational complexity, we resort to simulations to compute the distribution $b_{\rm out}^{(\ell)}(z)$ at the input of the decoder. A closed-form expression for 
$b_{\rm out}^{(\ell)}(z)$ is provided in~\cite{MaFeAbFrRa10} for $N=2$ and can be extended directly to scenarios with a generic value of $N$.

After a fixed number of internal message passing decoding operations,\footnote{In our numerical results with SCCCs, the number of internal iterations between convolutional decoders employing the BCJR algorithm is set to 10.} the extrinsic information sequence is extracted from the soft-output information sequence at the output of the decoder and the output SNR, denoted as  ${\rm SNR}_{\rm out}^{(\ell)}$, is evaluated. For a fixed value of the channel SNR, the above steps allow to numerically determine the $N$-dimensional input-output characteristic function $Z$ such that:
\begin{equation}
{\rm SNR}_{\rm out}^{(\ell)}=Z \left({\rm SNR}_{\rm in}^{(1)},\ldots,{\rm SNR}_{\rm in}^{(\ell-1)},{\rm SNR}_{\rm in}^{(\ell+1)},\ldots,{\rm SNR}_{\rm in}^{(N)},\gamma_\ell\right) \label{Z_LDPC_formula} .
\end{equation}

As previously shown, the component decoder can now be analyzed through a classical density evolution approach~\cite{RiUr01}, the only difference being the fact that the messages at the input of the decoder associated with the information bits need to be modified in order to model the presence of a priori information. In particular, in the iterative decoding procedure the a priori information from the other decoder is added to the channel information at the input of the information bits of the decoder. From a message density viewpoint, this corresponds to convolving the a priori message PDF $b_{\rm out}^{(\ell)}(z)$ by the Gaussian channel message PDF $\Gamma_{\rm ch}(z)$:
\begin{equation} \label{eq:modified}
m^{(\ell)}(z) = \Gamma_{\rm ch}(z) \otimes b_{\rm out}^{(\ell)}(z)
\end{equation}
where $\otimes$ denotes the convolution operator. However, one should note that this operation is performed \emph{only} in correspondence with the information bits, since the correlation model applies only to those bits. At this point, the density evolution procedure can be implemented in the classical way, by iterating the concatenated decoder or the sum-product algorithm for a fixed number of iterations. Note that the PDF at the input of the decoder is no longer exactly Gaussian, due to the transformation (\ref{eqn:n8}). However, the shape of $m^{(\ell)}(z)$ is similar to that of a Gaussian PDF (see, e.g.,~\cite{MaFeAbFrRa10}) and, therefore, one can conclude that the proposed EXIT chart-based approach is still accurate, although not exact. Numerical results, not reported here for lack of 
space, confirm this statement.

The characteristic values $\lambda_{\rm bal}$ and $\lambda_{\rm unb}$ can be obtained from the EXIT surface (\ref{Z_LDPC_formula}) in the following way. In the \emph{unbalanced} case, the value of the asymptote of the achievable region can be computed by, first, assuming that the a priori information sequences coming from the other decoders are characterized by sufficiently large SNRs, assumed to be equal, i.e., \mbox{${\rm SNR}_{\rm in}^{(m)}={\rm SNR}_{\rm in}\gg0$} \mbox{($m=1,\ldots,N$; $m\neq\ell$)}, and, then, finding the value of $\lambda_\ell$ (or, equivalently, $\gamma_\ell$) for which there is decoding convergence, e.g., ${\rm SNR}_{\rm out}\gg{\rm SNR}_{\rm in}$.

In the \emph{balanced} case, the $N$ channels are characterized by the same SNR $\gamma_\ell=\gamma_{\rm bal}$, \mbox{$\ell=1,\ldots,N$} (balanced channels). We can thus analyze the joint decoding convergence by drawing, for a given value of $\gamma$, the $Z$ hypersurface and its inverse, with respect to one of the inputs, $Z^{-1}$. Moreover, since the decoder is operating in a region characterized by the same SNR for all channels, it is reasonable to assume the same a priori SNR from all the sources, i.e., \mbox{${\rm SNR}_{\rm in}^{(m)}={\rm SNR}_{\rm in}$} \mbox{($m=1,\ldots,N$; $m\neq\ell$)}. Therefore, the analysis, carried out in~\cite{AbFeMaFrRa09,MaFeAbFrRa10} for a scenario with two sources, can be applied here, by considering the curve $Z_{\rm bal}=Z({\rm SNR}_{\rm in},\gamma)$ and characterizing the decoding convergence during external iterations (between the component decoders associated with the $n$ sources): the farther the curves, the faster the joint decoding convergence to  (theoretically) 
zero BER. When 
the two 
curves touch each other, then global decoding convergence 
is not achieved and the 
BER is bounded away from zero. The value of $\gamma_{\rm bal}$ (or, equivalently, of $\lambda_{\rm bal}$) is the minimum for which convergence is guaranteed.

\section{Performance Evaluation} \label{sec:performance}
We now evaluate the performance, in terms of achievable rate and bit error rate (BER), considering various pragmatic SCCC schemes. In particular, we assume that the transmitters use identical rate-$1/2$ linear codes, consisting of the cascade of an outer convolutional code (CC), a bit interleaver, and an inner CC~\cite{BeDiMoPo98b}. This code structure has been shown to guarantee a very good performance in a classical AWGN scenario~\cite{BeMo96,BeDiMoPo98b}. Moreover, as already observed in~\cite{GaZhZh07}, our results confirm that, for \emph{balanced} channels, SCCCs designed for the single-user AWGN scenario (i.e., without a priori information) work well also in the presence of correlation (i.e., with a priori information in a JCD scheme). However, the performance is still far from the capacity limit and further optimization is needed. On the other hand, in the \emph{unbalanced} case, i.e., when the SNRs in the channels are unequal, different channel coding strategies may entail better performance. 
Therefore, the design of good SCCCs, which allow to approach the theoretical performance limits, is of interest.

A possible example of an SCCC which works very well in the unbalanced regime was originally presented in~\cite{AbFeMa11} (on the basis of the guidelines given in~\cite{AbFe11}) and is characterized by the following generators:
\begin{eqnarray*}
G_{\rm inner}(D) &=& \left[\frac{1+D^2}{1+D+D^2+D^3}\right] \\
G_{\rm outer}(D) &=& \left[\frac{1+D^2}{1+D+D^2} \hspace*{5mm} \frac{1}{1+D+D^2}\right] .
\end{eqnarray*}
The inner and outer codes have rates $r_{\rm inner}=1$ and $r_{\rm outer}=1/2$, respectively. In the following, this SCCC will be denoted as SCCC$_1$. 

On the other hand, we denote as SCCC$_2$ the SCCC optimized for single-user AWGN scenarios with the following generators (as considered in~\cite{BeDiMoPo98b}):
\begin{eqnarray*}
G_{\rm inner}(D) &=& \left[1 \hspace*{5mm} \frac{1+D+D^2+D^3}{1+D^2+D^3}\right] \\
G_{\rm outer}(D) &=& \left[1+D^2+D^3 \hspace*{5mm} 1+D+D^2+D^3\right] .
\end{eqnarray*}
In this case, an overall 1/2 code rate is obtained via classical puncturing, by selecting coded bits alternately from the two component encoders with the following respective puncturing  matrices:
$$
P_{\rm inner} = \left[
\begin{array}{ccc}
        1 & 1 & 0 \\
        1 & 1 & 0
      \end{array}
      \right]
\hspace*{10mm}
P_{\rm outer} = \left[
\begin{array}{cc}
        1 & 1 \\
        1 & 0
      \end{array}
      \right].
$$
The inner and outer punctured CCs of SCCC$_2$ have thus rates $r_{\rm inner}=3/4$ and $r_{\rm outer}=2/3$, respectively.

We remark that although the original SCCC in~\cite{BeDiMoPo98b} has a rate 1/4 and operates with a 0.7~dB gap from capacity, other results, not shown here for lack of space, show that also the punctured rate-1/2 SCCC considered here (i.e., SCCC$_2$) guarantees a similar performance in a single-user AWGN scenario. It is worth noting that, unlike SCCC$_2$ (optimized for transmission over a single-user AWGN channel~\cite{BeDiMoPo98b}), SCCC$_1$ requires the use of an outer \emph{recursive} non-systematic CC. Moreover, observe that the inner code of SCCC$_1$ has rate 1 and, therefore, no puncturing is needed.

Finally, for comparison purposes we consider another SCCC, denoted as SCCC$_3$, which is optimized for a two-source scenario in the presence of a priori information (i.e., an unbalanced scenario), but it has very poor performance in the balanced regime~\cite{AbFeMa11}. The component CCs of SCCC$_3$ have the following generators:
\begin{eqnarray*}
G_{\rm inner}(D) &=& \left[1 \hspace*{5mm} \frac{1+D^2+D^3}{1+D+D^2+D^3}\right] \\
G_{\rm outer}(D) &=&  \left[\frac{1+D^2}{1+D+D^2}\right].
\end{eqnarray*}
The outer and inner CCs have rates $r_{\rm inner}=1/2$ and $r_{\rm outer}=1$, respectively. 

The need for a \emph{recursive outer CC} has been highlighted in~\cite{AbFe11}, where a BER-based optimization approach to design SCCCs in a single-source scenario with a priori information is presented. In fact, the scenario with a priori information considered in~\cite{AbFe11} is equivalent to the unbalanced case of interest in this paper, since a priori information is obtained from the correlation among sources. The results in~\cite{AbFe11} show that for sufficiently reliable a priori information and large code memory, a recursive outer CC is needed to optimize the performance (in~\cite{AbFe11}, an exhaustive search is performed by fixing the SCCC's overall code rate to 1/2).

We do not explicitly consider LDPC codes. In fact, classical systematic LDPC codes have similar performance to that of SCCC$_2$~\cite{AbFeMa11}, since they are optimized for a single AWGN channel. In the considered orthogonal multiple access schemes with correlated sources, proper LDPC code design is needed. For instance, in~\cite{YePfNa10,YePfNa12} the authors determine optimized node distributions of LDPC codes for a two-source scenario, together with a spatially-coupled graph-based detector. In particular, the LDPC code proposed in~\cite{YePfNa12} performs very well in all sub regions (balanced and unbalanced) of the achievable region and is thus referred to as ``universal.'' However, the generalization of this LDPC code design approach (and associated decoding strategy) to a scenario with $N>2$ represents an interesting research direction.

\subsection{EXIT Chart-based Analysis} \label{subsec:exit} 
We start analyzing the convergence of the considered codes by means of the EXIT chart-based method proposed in Section~\ref{sec:exit}. In particular, we first focus on the convergence of the codes denoted as SCCC$_1$ and SCCC$_3$ for large numbers of sources. We use the notation introduced in~\cite[Fig.~3]{AbFeMaFrRa09}, where $Z_1$ and $Z_2$ are the transfer functions of the outer and inner CCs, respectively. ${\rm SNR}_{\rm e,in}$ and ${\rm SNR}_{\rm e,out}$ denote, respectively, the SNRs of input and output extrinsic a priori information: the \emph{input} a priori information comes from the other $N-1$ decoders (relative to correlated sources), whereas the \emph{output} a priori information is desired to the other $N-1$ decoders. According to~\cite[Fig.~3]{AbFeMaFrRa09}, the input a priori information is fed only to the input of the inner decoder of each SCCC decoder.

\begin{figure}
\begin{center}
\includegraphics[width=0.48\textwidth]{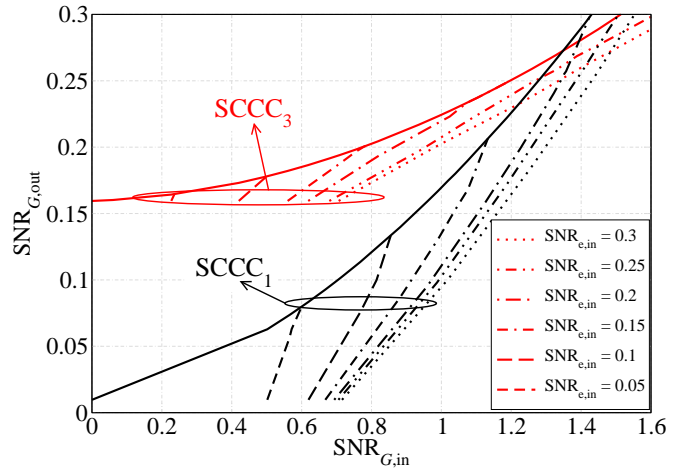}
\caption{EXIT charts of the component CCs for SCCC$_1$ and SCCC$_3$ and various values of the SNR of input extrinsic a priori information. The number of sources is equal to $N=50$, the channel SNR is $\gamma=-5.2$~dB, and $\rho=0.95$. Solid lines correspond to the $Z_2$ curve, whereas the others correspond to $Z_1^{-1}$.}
\label{fig:exit_SCCC}
\end{center}
\end{figure}

\begin{figure}
\begin{center}
\includegraphics[width=0.48\textwidth]{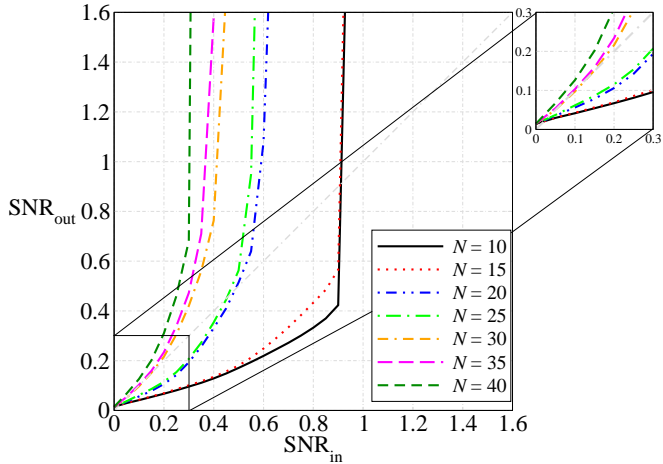}
\caption{EXIT charts of SCCC$_1$ in the balanced regime for $\gamma=-5.2$~dB, $\rho=0.95$, and various values of $N$.}%
\label{fig:exit_balanced}
\end{center}
\end{figure}

In Fig.~\ref{fig:exit_SCCC}, we show the EXIT charts of the component CCs for various values of the SNR of input extrinsic a priori information. The number of sources is equal to $N=50$, $\rho=0.95$, and the channel SNR is $\gamma=-5.2$~dB (which is the asymptotic value of the threshold achievable by both SCCCs for large numbers of sources, as it will be more clearly shown in Fig.~\ref{fig:exit_balanced}). In the figure, the subscript $G$ corresponds to the considered CC, i.e., $G=1$ for the inner code, $G=2$ for the outer code.
As in classical EXIT chart-based analysis, convergence is guaranteed if the tunnel between the transfer functions of the component CCs is open. This is the case if the SNR of the extrinsic a priori information (i.e., ${\rm SNR}_{\rm e,in}$) is sufficiently high. Since a large a priori information corresponds to the unbalanced scenario of interest in this paper, we can conclude that both codes converge in this scenario. When the SNR of the a priori information is sufficiently small (corresponding to the balanced case of interest in this paper), instead, convergence is guaranteed only for SCCC$_1$ and not SCCC$_3$. In fact, for SCCC$_1$ ${\rm SNR}_{\rm e,out}=0.08$ is achieved for an input value ${\rm SNR}_{\rm e,in}=0.05$ and, therefore, the extrinsic information increases with code iterations. With SCCC$_3$, instead, one obtains ${\rm SNR}_{\rm e,out}=0.038$ for ${\rm SNR}_{\rm e,in}=0.05$. This means that the extrinsic information decreases with code iterations and no 
convergence can be achieved. This behavior can be 
explained by observing that SCCC$_1$ has a powerful half-rate outer CC which can perform well also in the absence of reliable a priori information. SCCC$_3$, instead, is characterized by a rate-1 outer code which is not reliable for limited a priori information (small values of the extrinsic SNR).

In Fig.~\ref{fig:exit_balanced}, we present the EXIT charts of SCCC$_1$ in a balanced regime. In particular, the value $\gamma=-5.2$~dB guarantees convergence in this regime. In all cases, $\rho=0.95$ and various values of $N$ are considered.
First, one should observe that all curves tend to infinity for sufficiently large values of the input SNR, meaning that there is convergence, in the unbalanced scenario, for all considered values of $N$. Moreover, for $N\geq30$ all EXIT curves are above the bisector line (see the zoom on the side) and, therefore, the tunnel opens. This confirms that both the balanced and the unbalanced cases converge to the same SNR value for sufficiently large values of $N$ (in this case $\gamma=-5.2$~dB).

\subsection{Achievable Rate Results} \label{subsec:rate}
We now present the achievable region for the considered SCCCs. The achievable rates are measured by simply extrapolating, from the EXIT charts, the lowest rate for which the tunnel opens. Our simulation results show a good agreement between the EXIT chart-based rate identification method and BER-based simulation results. In the latter case, the rate is determined, through the formula $\lambda=0.5\log_2(1+\gamma)$, from the SNR value for which BER$=10^{-5}$. Although the used capacity formula is exact only for Gaussian inputs, it is approximately valid also for BPSK modulation in the low-rate regime of interest here. These results are partly presented in~\cite{AbFeMaFrRa09} and the others are not shown here for conciseness.

\begin{figure}
\begin{center}
\begin{tabular}{c}
\includegraphics[width=0.48\textwidth]{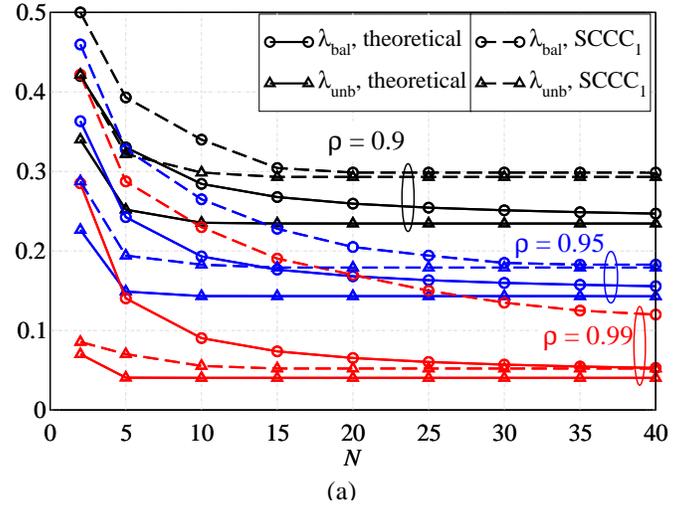} \\
(a)\\
\includegraphics[width=0.48\textwidth]{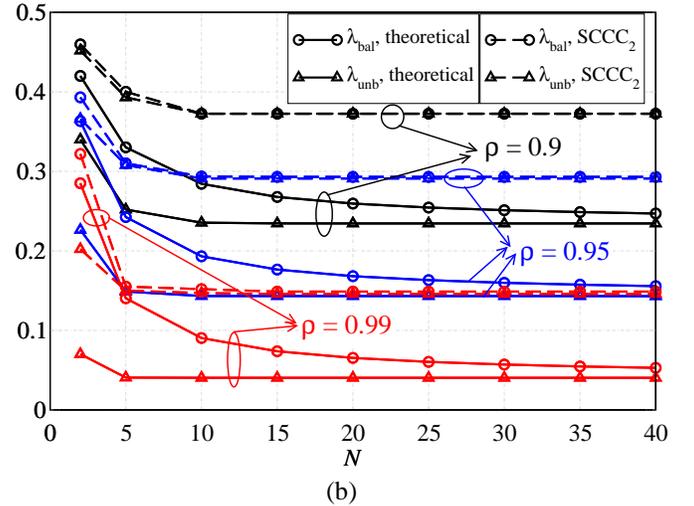}\\
(b)
\end{tabular}
\caption{$\lambda_{\rm bal}$ and $\lambda_{\rm unb}$ of the considered SCCC$_1$ (case~(a)) and SCCC$_2$ (case~(b)), as functions on $N$, for various values of $\rho$. Theoretical limits are also reported as performance benchmarks.}%
\label{fig:FR_codici}
\end{center}
\end{figure}

In Fig.~\ref{fig:FR_codici}, the values of $\lambda_{\rm bal}$ and $\lambda_{\rm unb}$, with SCCC$_1$ (case~(a)) and SCCC$_2$ (case~(b)), are shown, as functions on $N$, for various values of $\rho$. Theoretical limits are also reported as performance benchmarks.
First, one should observe that in the unbalanced regime SCCC$_1$ always outperforms SCCC$_2$: this is expected, as SCCC$_1$ is optimized for these settings. On the other hand, SCCC$_2$ has a better performance in the balanced case for small values of $N$ (e.g., $N\leq5$), since the balanced case is ``more similar'' to a single-user AWGN scenario. However, when the number of sources increases, SCCC$_2$ achieves the performance limits of the unbalanced case (as predicted by our theoretical framework), but this asymptotic limit is lower than that of SCCC$_1$. Therefore, SCCC$_1$ becomes more effective. These results confirm the predictions of our theoretical framework and lead to pragmatic channel coding schemes that can be directly applied (with good performance) to large scale scenarios.

\begin{figure}
\centering
\includegraphics[clip,width=0.48\textwidth]{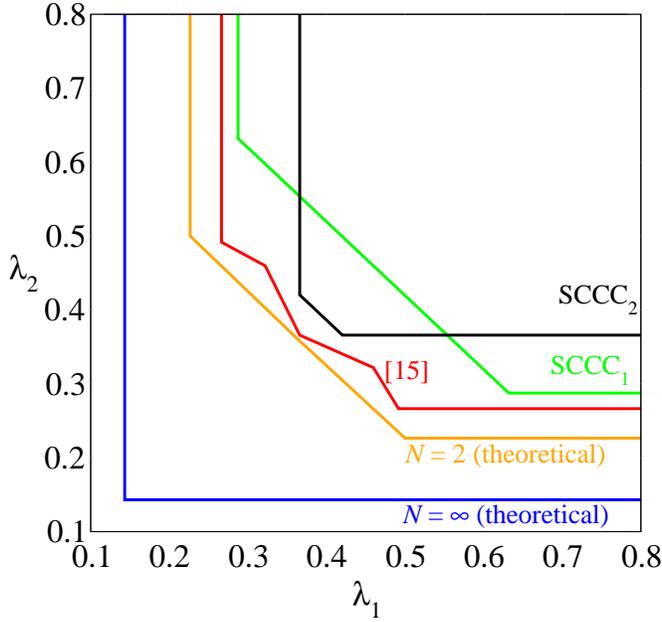}
\caption{Two-dimensional ($N=2$) region for the following coded schemes: SCCC$_1$, SCCC$_2$, and the LDPC code-based proposed in~\cite{YePfNa12}.}
\label{fig:pfister}
\end{figure}

For $N=2$ sources, it is of interest to compare the performance of the proposed ``off-the-shelf'' SCCCs with that of the ``universal'' LDPC coded scheme proposed in~\cite{YePfNa10,YePfNa12}. In Fig.~\ref{fig:pfister}, we directly compare the two-dimensional achievable regions for the following three coded schemes: SCCC$_1$-, SCCC$_2$-, and LDPC code-based, the latter from~\cite{YePfNa12}.
One can observe that the coded scheme in~\cite{YePfNa12} outperforms our schemes. In fact, it is optimally designed for all operational zones (both balanced and unbalanced), with a spatially coupled decoding approach operating on the overall joint graph obtained from the two component Tanner graphs. The proposed approach, instead, is based on the idea of passing messages between ``standard'' pragmatic component decoders. Despite this inherent suboptimality, one can observe that SCCC$_1$, which is optimized for the unbalanced case, allows to approach the performance of the scheme in~\cite{YePfNa12} in the unbalanced zone, with a standard decoding structure. Similarly, the scheme based on SCCC$_2$ (optimized for the balanced regime) approaches the performance of the scheme in~\cite{YePfNa12} in the balanced zone. Building upon the proposed pragmatic approach, our SCCC schemes can be implemented and perform well with relatively short codewords, whereas, to the best of our knowledge, we are not aware of a finite-
length parity-check matrix for the LDPC code in the scheme in~\cite{YePfNa12}---the results presented in~\cite{YePfNa12} are based on a belief propagation-based analysis for infinite codeword length.

\subsection{BER Results} \label{subsec:ber}
\begin{figure}
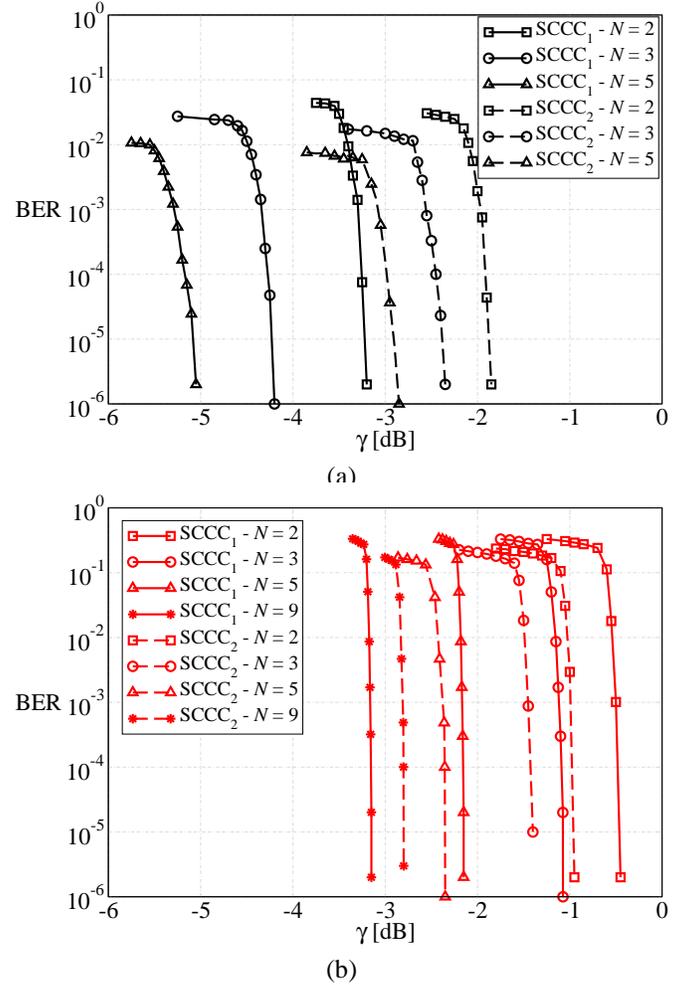

\begin{center}
\begin{tabular}{c}
\includegraphics[width=0.48\textwidth]{./images/ber_unbalanced.eps} \\
(a)\\
\includegraphics[width=0.48\textwidth]{./images/ber_balanced.eps}\\
(b)
\end{tabular}
\caption{BER, as a function of the SNR, for different coding schemes, number of sources, and $\rho=0.95$. The \emph{unbalanced} case is considered in~(a), whereas the \emph{balanced} case is considered in~(b).}%
\label{fig:ber_unbalanced}
\end{center}
\end{figure}

We now investigate the performance, in terms of BER, of some of the proposed coding schemes (namely, SCCC$_1$- and SCCC$_2$-based), considering various values of the number of sources.\footnote{The performance of the SCCC$_3$-based scheme is not shown since it is much worse than that of the other schemes in the balanced case and similar to that of the SCCC$_1$-based scheme in the unbalanced case.} All simulations have been performed using an information word length equal to 50000. In Fig.~\ref{fig:ber_unbalanced}, the BERs of the proposed SCCC-based schemes, in both (a) \emph{unbalanced} and (b) \emph{balanced} scenarios, is analyzed, considering various values of $N$.
In the unbalanced case, one can observe that the SCCC$_1$-based scheme, whose design is optimized for this scenario, significantly outperforms the SCCC$_2$-one scheme. As expected, the SCCC$_2$-based scheme has very poor performance in the unbalanced case, as it is optimized for a single-source scenario and cannot properly exploit the a priori information coming from other decoders, even when this information is very reliable. In the balanced case, instead, one should note that the SCCC$_2$-based scheme outperforms the SCCC$_1$-based scheme for small values of $N$, whereas the latter scheme becomes preferable for larger numbers of sources (e.g., $N=9$). It can be observed that, for increasing values of $N$, the performance improvement of the SCCC$_1$-based scheme, with respect to the SCCC$_2$-based one, increases. This is in agreement with the convergence behavior of the (theoretical) achievable region analyzed in Section~\ref{sec:fr}, according to which a good coding scheme for the unbalanced case, for 
increasing values of $N$, guarantees a good performance in the balanced case as well.  This is also in agreement with the results in Fig.~\ref{fig:FR_codici} for the achievable rates.
 
On the basis of the theoretical considerations in Section~\ref{sec:fr} and the above presented results, practical channel code design guidelines can be summarized as follows. If a channel code is optimized in the unbalanced scenario, then it \emph{may} reach, for increasing values of $N$, the same asymptotic limit also in the balanced case. However, the effective behavior in the balanced scenario depends on the specific code design. For instance, as shown in Fig.~\ref{fig:exit_SCCC} through an EXIT chart-based analysis, both SCCC$_1$ and SCCC$_3$ perform well in the unbalanced case, but in the balanced scenario SCCC$_3$ does not converge, while SCCC$_1$ does, as confirmed by the BER-based results in Fig.~\ref{fig:ber_unbalanced}~(b). If, on the other hand, the channel code has limited convergence in the unbalanced case, then the asymptotic limit will never be reached in the balanced case. In Fig.~\ref{fig:ber_unbalanced}~(a), SCCC$_1$ is shown to outperform SCCC$_2$ in the unbalanced scenario for all 
considered values of $N$, and the relative gap increases for increasing values of $N$. This indicates that the asymptotic limit of SCCC$_1$ in the unbalanced case is closer to the theoretical limit than that of SCCC$_2$, suggesting that SCCC$_1$ might asymptotically outperform SCCC$_2$ also in the balanced case: this happens for sufficiently large values of $N$, i.e., $N\geq9$. However, for smaller values of $N$ (i.e., $N<9$), SCCC$_2$ outperforms SCCC$_1$ in the balanced case.

\begin{figure}
\centering
\includegraphics[width=0.48\textwidth]{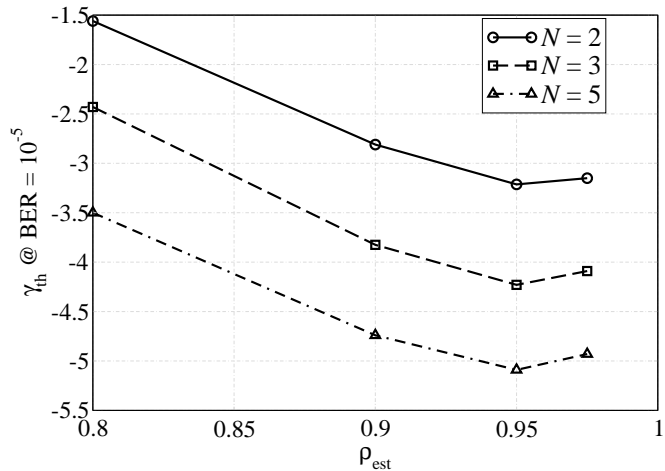}
\caption{SNR threshold, as a function of $\rho_{\rm est}$, in the unbalanced case and with $\rho=0.95$. Various values of $N$ are considered, and in all cases the SCCC$_1$-based scheme is used.}
\label{fig:mismatch}
\end{figure}

Finally, it is of interest to understand what is the impact of the number of sources when the correlation model (i.e., the value of the parameter $\rho$) is not perfectly known at the AP. To this end, we investigate a mismatched scenario in which, while the transmitted data are generated according to the correlation model introduced in Section~\ref{sec:scenario} with parameter $\rho$, the receiver uses an estimate $\rho_{\rm est}$ of the true correlation parameter $\rho$. The performance is evaluated in terms of the minimum SNR required to guarantee a target BER equal to $10^{-5}$. In Fig.~\ref{fig:mismatch}, the threshold SNR (denoted as $\gamma_{\rm th}$) is shown, as a function of $\rho_{\rm est}$, in the unbalanced case and with $\rho=0.95$. Various values of $N$ are considered, and in all cases the SCCC$_1$-based scheme is used.
As expected, the threshold SNR is lowest for $\rho=\rho_{\rm est}$, i.e., when exact correlation knowledge is available at the receiver. As already observed in~\cite{AbFe11} for $N=2$, SCCC$_1$ guarantees a limited penalty even in the presence of a moderate reliability estimation error. In other words, SCCC$_1$-based schemes are robust against estimation errors. Moreover, the number of sources has no impact on the ``shape'' of the threshold BER curve.

\section{Concluding Remarks} \label{sec:concluding}
In this paper, we have considered orthogonal multiple access schemes with a generic number of correlated sources, where the correlation is exploited at the AP. In particular, each source uses an ``off-the shelf'' channel code to transmit, through an AWGN channel, its information to the AP, where component decoders, associated with the sources, iteratively exchange soft information by taking into account the correlation. In the presence of $N$ sources, we have first characterized the multi-dimensional achievable region, defined as the ensemble of the channel parameter $N$-tuples where arbitrarily small probability of error is achievable. Our results show that, for asymptotically large values of the number of sources, the achievable region approaches a translated hyperoctant. We also discussed the speed of convergence of the achievable region to this translated hyperoctant. Then, on the basis of an EXIT chart-based approach, we computed the main (i.e., in unbalanced and balanced scenarios) 
parameters of the achievable region for a few representative pragmatic SCCCs. Our results with these pragmatic coding schemes confirm the predictions of our information-theoretic framework: in particular, a good coding scheme for the unbalanced scenario tends, for large values of $N$, to achieve the same asymptotic performance in the balanced case as well. We also presented BER-based results of our coding schemes, thus showing that they can be effectively implemented for finite-length codewords.

\section*{Acknowledgments}
The authors would like to thank the Associate Editor and the anonymous Reviewers for their insightful suggestions which helped to improve the quality of this paper.

\bibliographystyle{IEEEtran}
\bibliography{references}

\end{document}